\documentclass[sigconf, nonacm, screen]{acmart}

\usepackage{include/packages}
\usepackage{include/commands}
\usepackage{include/shorthands}

\begin{document}

\title{\ours{}: Next-Level Distributed Machine Learning Simulations via High-Fidelity GPU and Infrastructure Modeling}

\author{%
William~Won$^{*1}$,\space{}
Jinsun~Yoo$^{*2}$,\space{}
Tuan~Ta$^{*1}$,\space{}
Moumita~Dey$^{*1}$,\space{}
Andy~Balogh$^{3}$,\space{}
Pradosh~Datta$^{3}$,\space{}
Furkan~Eris$^{1}$,\space{}
Conor~Green$^{1\text{~}4}$,\space{}
Winston~Liu$^{3}$,\space{}
Changhai~Man$^{2}$,\space{}
Kingshuk~Mandal$^{3}$,\space{}
Amos~Rai$^{3}$,\space{}
Vinay~Ramakrishnaiah$^{1}$,\space{}
Ruchi~Shah$^{1}$,\space{}
David~Sidler$^{1}$,\space{}
Harsh~Sikhwal$^{3}$,\space{}
Hanjiang~Wu$^{2}$,\space{}
Tushar~Krishna$^{\text{†}2}$,\space{}
and~Bradford~M.~Beckmann$^{\text{†}1}$
}
\affiliation{%
\vspace{0.5em}
\institution{%
    $^{1}$AMD Research and Advanced Development\space{}\space{}
    $^{2}$Georgia Institute of Technology\space{}\space{}
    $^{3}$Keysight\space{}\space{}
    $^4$Purdue University
}
\city{}
\country{}
\vspace{1em}
}
\thanks{%
    * These authors contributed equally to this work.\space{}\space{}
    † Equal advising.
    
    Correspondence to: 
    William Won <William.Won@amd.com>, 
    Tushar Krishna <tushar@ece.gatech.edu>, 
    Bradford M. Beckmann <Brad.Beckmann@amd.com>.
}

\begin{abstract}

Distributed machine learning~(ML) is a key paradigm for today's large-scale artificial intelligence applications.
As model inference arises as an important use case, faithful modeling of latency-sensitive collective communication has never been more important.
Capturing the device architecture and modeling control and data paths at high fidelity is therefore a necessity today.
Having a common, detailed representation for distributed ML infrastructure is also crucial.
We revisit the promising open-source, community-driven simulator: ASTRA-sim.
In this work, we identify limitations of the current ASTRA-sim simulator and augment it with new features.
To this end, we enable fine-grained, high-fidelity simulation with a standardized infrastructure representation, opening new design space exploration opportunities.
We propose the simulation at cache-line-sized load-store granularity, with a detailed graphics processing unit~(GPU) execution model, to balance simulation scalability and fidelity.
We also introduce InfraGraph, a standardized representation to capture distributed ML network infrastructure in detail.
Using the updated ASTRA-sim 3.0 simulator, we showcase interesting design space explorations for designing optimized collective algorithms, network requirements, and GPU architectures.

\end{abstract}

\maketitle
\renewcommand*{\shortauthors}{Won et al.}

\section{Introduction}\label{sec:Introduction}

Today, the importance of foundation artificial intelligence~(AI) models needs no emphasis.
Generative AI models, including large language models~(LLMs), have been adopted in numerous domains and have been scaling at an unprecedented rate.
On the training front, the compute requirements for frontier models have been increasing 4--5$\times$ per year on average~\cite{epoch2026trends}, with LLMs comprising trillions of parameters~\cite{fedus2022switchtransformers, deepseekv4}.
Inference outpaces the already massive training requirements, even taking 80--90\% of energy usage and cost throughout the model life cycle~\cite{caravaca2025frompromptstopowera, patel2022therealpriceofai}.
To accommodate such scales, the use of distributed machine learning~(ML) is necessitated; state-of-the-art data centers contain tens to even hundreds of thousands of graphics processing units~(GPUs)~\cite{xaiColossus, metainfra, openai2026IntroducingStargateNorway}.

Large-scale distributed ML requires massive resources---not only to train models, but also to service the increasing demands for inference~\cite{mit2026energy, chen2025electricitydemandandgridimpactsofaidatacenters}.
Considering these needs, optimizing for the performance and cost of distributed ML is crucial.
However, the co-design space of distributed ML is vast and complex~\cite{won2023astrasim2.0}.
It comprises multiple design stacks---AI models, parallelization strategies, GPU architectures, and communications between GPUs---and the interplay across them~\cite{rashidi2020astrasim}.
Navigating this intertwined design space requires a systematic and methodical approach.
Therefore, implementing an infrastructure to capture this full-stack design space of distributed ML remains pivotal research.

We revisit an existing solution: \astrasim{}~\cite{rashidi2020astrasim, won2023astrasim2.0}.
It is an event-driven, full-stack distributed ML simulator.
We target \astrasim{} since it is an open-source, community-driven effort and has been adopted by multiple works in the field~\cite{cho2024llmservingsim, cho2026llmservingsim2.0, man2025stage, yoo2024StandardizedRepresentationDeepa, yoo2025EasyRealisticNetwork}.
\astrasimone{} first introduced this simulation infrastructure, but supported only a limited set of training workloads and network topologies~\cite{rashidi2020astrasim}.
\astrasimtwo{}, a subsequent update, identified such issues and improved the simulator~\cite{won2023astrasim2.0}.
Notably, \astrasimtwo{} incorporates graph-based workload representation~\cite{sridharan2023chakra} and extends the scope of network topologies.
Despite greater flexibility, considering today's challenges associated with distributed ML, \astrasimtwo{} still comes with limitations that were overlooked at the time.
In this work, we aim to further identify such restrictions and implement new features in the \astrasim{} framework.
To this end, we target capturing the broader design space with improved accuracy.

Firstly, we point out the limitations of \astrasim{}'s collective communication modeling capabilities.
For both training and inference jobs, collective communication is the dominant operation~\cite{wang2025SystematicCharacterizationLLM, chen2025EfficientReliableObservable}, yielding multiple fast-paced research efforts.
They are spread across new collective algorithm designs~\cite{cho2023LogicalPhysicalTopologyAware, laskar2024EnhancingCollectiveCommunication, ma2021paard} and automated collective algorithm synthesis~\cite{zhao2026forestcoll, won2024tacos, wang2019blinka, shah2023taccl, cai2021SynthesizingOptimalCollective, huang2021codesign, kim2024tccl, liu2024RethinkingMachineLearning}, to list a few.
However, \astrasim{} only supports a limited set of predefined, textbook collective algorithms: ring~\cite{thakur2005mpich}, all-pairs (i.e., direct)~\cite{rashidi2020astrasim}, double binary tree~\cite{sylvainjeaugey2019dbt}, and recursive halving-doubling~\cite{thakur2005mpich}.
Although collective communication is on the critical path, \astrasim{} cannot capture the latest research endeavors without manual implementation.

\insertFigureWide{AstraSimOverview}{1}{0}{0}{
Overview of the \ours{} infrastructure.
New and improved components are marked with bold red borders.
}

Secondly, \astrasim{} lacks device modeling.
It completely overlooks the control path and resource contention.
Capturing such effects is especially pivotal today, when inference tasks have become as important, if not more so, as training executions.
Inference jobs are latency-sensitive~\cite{zhao2025MLInferenceScheduling}, making accurate latency modeling essential.
This requires faithful modeling of fine-grained architectural details.
For example, assume a GPU wants to write a chunk of data to a remote GPU.
This actually consists of multiple steps:
(\rom{1})~a compute unit\footnote{
We use OpenCL/AMD and NVIDIA terminology interchangeably, though the former is preferred (e.g., compute unit vs. streaming multiprocessor, workgroup vs. threadblock, wavefront vs. warp).
}~(CU)
loads cache-line-sized (e.g., 128\,B~\cite{AMDCDNA4, nvidiaPascalTuningGuide}) data from the local high-bandwidth memory~(HBM) to its register file
(\rom{2})~the CU then writes the data to the input/output~(I/O) port of the socket
(\rom{3})~the network transfers the cache-line-sized data to the remote GPU's I/O port
(\rom{4})~the remote GPU writes the received data to the destination HBM.
These steps are repeated at cache-line granularity until all data chunks are written to the destination.
However, the current \astrasim{} simulates this as a single network transfer, blocking accurate latency estimation.
The problem is exacerbated by other important missing control path modeling, including memory fences, intra-GPU barriers, and inter-GPU semaphores.
Multiple GPU kernels fight for the limited CU resources, affecting the latency as well.
All such effects must be captured for correct latency modeling.

To properly simulate such fine-grained operations, the network model should support network-on-chip~(NoC)-level details.
However, the current \astrasim{} infrastructure only assumes coarse-grained, inter-GPU communications.
It neglects on-chip transfers such as a CU loading cache-line-sized data from the local HBM channel.
The network simulation backend should also be upgraded to support the modeling of the architectural details of a GPU socket.

Finally, the community requires a common, detailed representation of distributed ML network infrastructure.
Currently, different network simulation backends all require unique input formats.
As different simulators capture infrastructure at distinct levels with different details, this fragmentation not only hinders accurate infrastructure modeling but also prohibits the community from exchanging their infrastructure details.

In this work, we introduce \ours{}: taking distributed ML simulations to the next level via high-fidelity modeling.
\autoref{fig:AstraSimOverview} summarizes the improved \ours{} simulation infrastructure.
\ours{} implements new features to eliminate the limitations and to faithfully model today's distributed ML systems.
\ours{} supports arbitrary customized collective algorithms through \mscclpp{} representations~\cite{shah2025mscclpp}.
\ours{} purposely represents workloads at \loadinst{}-\storeinst{} granularity, capturing fine-grained details while remaining scalable.
A new \gpumodel{} simulates them at cache-line-sized granularity with CU, workgroup, and wavefront-level modeling, just as real GPUs do.
Finally, we build \infragraph{}, a graph-based backend-agnostic representation of the physical system.
Users can exchange and reuse a single \infragraph{} description across network backends in \ours{}.

\section{Background}\label{sec:Background}

\subsection{\astrasim{}}

\astrasim{}~\cite{rashidi2020astrasim, won2023astrasim2.0} is an open-source distributed ML simulator.
The high-level architecture is summarized in~\autoref{fig:AstraSimOverview}.
It consists of three building-block layers: workload, system, and network.
The workload layer captures the design space of the target model and its parallelization, whereas the network layer simulates the network transfers.
The system layer intermediates between the two layers, for example, estimating compute kernel runtime or breaking down a collective communication kernel into chunk-granularity transfers using predefined collective algorithms.

\astrasimone{} is the introductory version but lacked flexibility in all three layers~\cite{rashidi2020astrasim}.
It only supported a predefined set of training workloads, collective algorithms, and network topologies (specifically, two-dimensional switch and torus) through Garnet, a network model originally developed in gem5~\cite{lowe-power2020gem5}.
Later, an application programming interface~(API) to plug in arbitrary network simulators was introduced, and ns-3~\cite{hendersonNetworkSimulationsNs3} was incorporated as another simulation backend~\cite{rashidi2020ns3}.
\astrasimtwo{} addressed some flexibility limitations~\cite{won2023astrasim2.0}.
For the workload layer, \astrasimtwo{} leverages MLCommons Chakra execution trace~(ET)~\cite{sridharan2023chakra} to support arbitrary workloads.
For the network layer, \astrasimtwo{} introduces a new $\alpha$--$\beta$-model-based~\cite{Hockney1994TheCC} \simple{} network simulator.\footnote{
We name it \simple{} in this work (previously called \texttt{Analytical}), following its resemblance to the \simple{} network simulator in gem5~\cite{Gem5InterconnectionNetwork}.
}

\subsection{Graphics Processing Unit}\label{sec:Background:GPUArchitecture}

We briefly introduce the background on GPUs, the dominant compute devices in distributed ML~\cite{googleWhatGPUIts, john2024performanceandpowera}.

\subsubsection{Programming Model}

A kernel is a function to be executed on a GPU, and comprises many threads.
Each thread executes the same instruction but with different data, making the GPU follow the single-instruction, multiple-data~(SIMD) paradigm.
Threads in a kernel are organized into workgroups~(\ie{} threadblocks), a composition of threads (\eg{} 1,024 threads).
All threads in a workgroup are mapped and executed on one CU~(\ie{} streaming multiprocessor~(SM)).
Within a workgroup, threads are further grouped into a smaller unit called wavefronts~(\ie{} warps), a lock-step execution unit within a CU (\eg{} 32 threads).
\autoref{fig:GpuArchOverview} depicts a GPU kernel with four workgroups, each containing two wavefronts.

\subsubsection{GPU Architecture}
GPU hardware consists of many CUs to execute multiple workgroups in parallel.
When a kernel is dispatched to the GPU, each workgroup is mapped onto a CU.
The CU then executes threads in the wavefront in lock-step.
\autoref{fig:GpuArchOverview} shows that the four workgroups in the kernel are being executed in parallel by four CUs.

\insertFigure{GpuArchOverview}{1}{0}{0}{
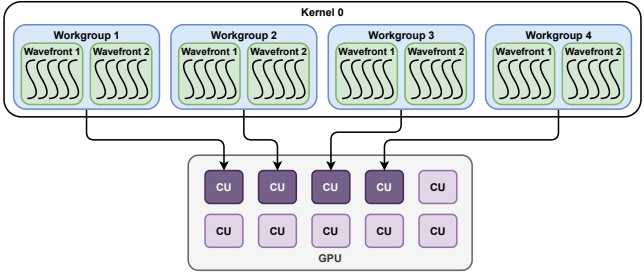
Abstract view of GPU program and architecture.
}

\subsection{Collective Communication}

The model and/or data are dispersed across many devices in distributed ML, requiring that the devices frequently synchronize to execute the full workload.
Such communication takes place in the form of collective communication patterns defined in the message passing interface~(MPI)~\cite{mpi2025doc}.

A collective algorithm defines how individual data chunks are transferred over a network topology to execute the collective patterns.
Textbook collective algorithms are well-defined, basic collective algorithms that are pervasively used in today's collective communication libraries~(CCLs).
Ring~\cite{thakur2005mpich}, all-pairs (i.e., direct)~\cite{rashidi2020astrasim}, double binary tree~\cite{sylvainjeaugey2019dbt}, and recursive halving-doubling~\cite{thakur2005mpich} are examples of predefined textbook collective algorithms used by CCLs today.

\subsection{\mscclpp{} Custom Collective Representation}\label{sec:Background:mscclpp}

As collective communication becomes a crucial part of distributed ML, many research efforts have been put into the domain; among them is the representation of collective algorithms.
Notably, \mscclang{} introduces a domain-specific language~(DSL) to represent arbitrary collective communication algorithms~\cite{cowan2023mscclang}.
Customized collective algorithms can be written in a Python-based DSL.
Later, \mscclpp{} was introduced to generalize the DSL and to represent a more flexible set of collective algorithms~\cite{shah2025mscclpp}.
For instance, \mscclpp{} supports the use of one-sided \texttt{put}/\texttt{get} operations, allows one workgroup to talk to multiple remote GPUs simultaneously, and captures control dependencies via barriers and semaphores.
\mscclpp{} compiles the DSL into a custom JavaScript object notation~(JSON) file, capturing workgroup-level operations of individual GPUs.
\autoref{fig:JsonSnippetExample} captures a simplified view of the JSON file that represents a collective operation of two GPUs.
Each GPU comprises two workgroups, each with three GPU operations.

\insertFigure{JsonSnippetExample}{0.9}{-0.5}{0}{
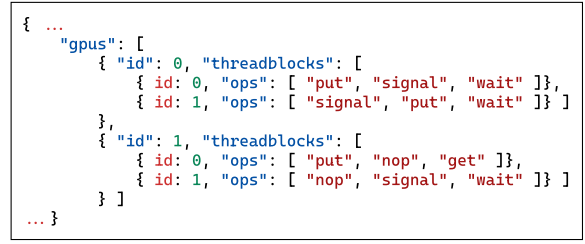
A simplified example of \mscclpp{} JSON collective algorithm representation.
}

\section{Motivation}\label{sec:Motivation}

Here, we identify the limitations of the current \astrasim{}.
We discuss why having such modeling capabilities is important for solving today's distributed ML challenges.

\subsection{Customized Collective Algorithms}

The current \astrasim{} infrastructure completely lacks customized collective algorithm modeling capabilities.
As \autoref{sec:Introduction} discusses, collective communication has become the major bottleneck in distributed ML.
For training tasks, collective sizes are often large, and the target is to optimize for throughput, while inference communication is smaller in size and latency-sensitive.
Such different natures yield distinct approaches and techniques for collective optimization.
To list a few, even for textbook collective algorithms, experts fine-tune the number of workgroups, the size of individual chunks, communication protocols, or communication primitives (e.g., two-sided vs. one-sided transfers).
Collective algorithm synthesizers automatically generate topology-aware collective algorithms.
As the system scales, fault-tolerant collective algorithm design also becomes an important research angle.
Ultimately, the existence and adoption of \mscclpp{} representation itself highlight the necessity to execute customized collective algorithms.

However, \astrasim{} infrastructure only supports a limited set of predefined textbook collective algorithms, with very limited design parameters (e.g., number of chunks per kernel).
\astrasim{} cannot simulate today's fast-changing collectives without extensive manual implementation.

\subsection{Fine-Grained GPU Modeling}

\insertFigure{MotivationLLSimple}{1}{0}{0}{
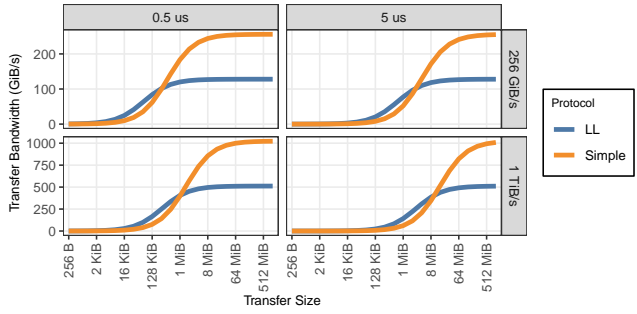
Analytical analysis of LL versus Simple transfer bandwidth, using different link latency and bandwidth.
}

The existing \astrasim{} infrastructure does not capture fine-grained device modeling.
With the massive adoption of AI workloads, the community is shifting gears from training to ML inference.
As inference workloads are highly latency-sensitive, correct latency modeling is essential.

Faithful latency modeling necessitates high-fidelity modeling of the data and control paths of the device.
As articulated in~\autoref{sec:Introduction}, even a simple network \texttt{put} operation actually involves multiple repetitive steps working at cache-line granularity:
(\rom{1})~after making sure the destination buffer is ready,
(\rom{2})~a CU reads cache-line-sized data to the register file, and
(\rom{3})~initiates the network transfer of the read value; then
(\rom{4})~the remote GPU stores the received data to the destination.
However, none of these fine-grained operations are modeled in \astrasim{}.
This is not to mention that multiple workgroups from different communication and compute kernels contend for the limited CU and memory resources, further impacting the overall latency.

\autoref{fig:MotivationLLSimple} is a quantitative motivation for why such accurate latency modeling is required.
Most CCLs in use today provide two communication protocols: low-latency~(LL) and Simple.
The Simple protocol can effectively leverage 100\% of network bandwidth but comes with the tradeoff of synchronization before and after the transfer.
The LL protocol eliminates the synchronization for latency optimization with the tradeoff of 50\% link bandwidth.
Intuitively, LL is preferred for small transfers, and Simple starts to outperform as the transfer size increases.
Using the analytical \alphabeta{} modeling with arbitrary latency and bandwidth values, we plotted the transfer bandwidth of the two protocols with different transfer sizes.
When we underestimate the transfer latency (as 0.5\,$\mu$s), the Simple protocol starts to outperform LL much faster.
For 256\,GiB/s link bandwidth, the Simple protocol outperformed LL at a 512\,KiB transfer size.
However, if we overestimate the link latency (as 5\,$\mu$s), the Simple protocol performance saturates at much larger transfer sizes; it only outperformed the LL protocol at a 2\,MiB transfer size.
As the link bandwidth increases to 1\,TiB/s, the gap is even larger: 4\,MiB versus 16\,MiB.
This simple analytical analysis underscores that improper latency modeling can lead to wrong design conclusions, which becomes especially crucial as the network becomes more performant.
High-fidelity latency modeling through fine-grained GPU modeling is therefore necessitated in the simulation infrastructure.

\subsection{Common Infrastructure Representation}

To accommodate different use cases (e.g., fidelity versus scalability), \astrasim{} extends to multiple network simulation backends via common network APIs.
However, users should define the physical system topology in a format specific to the target network backend.
Examples include routing configurations, the duration of a message, or whether to model congestion at a given point.
Through our experience using \astrasim{}, we found the lack of a global representation format to be problematic.

Firstly, any physical system description is tied to a specific network backend.
This fragmented approach forces users to manually rewrite infrastructure descriptions across network backends, not to mention the time to understand the semantics of the format, which is not intuitive and requires per-backend knowledge.
Secondly, it hinders the community from sharing and exchanging the exact infrastructure details.
Existing research often describes physical systems with high-level verbiage such as ``two-tiered fat-tree topology" or through simple figures.
However, this may not necessarily capture the full detail of the topology, making it vague when translating into a description that \astrasim{} backends can understand.
If the community had a common, standardized means to represent their physical topology information in a neutral format, it could facilitate the community in sharing infrastructure details and reproducing the simulation results.
We envision a potential beyond \astrasim{} in standardizing the infrastructure representation.

In short, the above motivates a backend-agnostic format that can represent arbitrary physical system topologies.
We motivate the need for a single format that, once defined, not only captures the fine-grained details of the network infrastructure, but also can be easily shared across the research community.
Such representation can be automatically translated to the backend-specific format for \astrasim{}.

\section{\ours{}}\label{sec:Contribution}

Here, we explain the new features added to \ours{}.
We also articulate their implementation in detail.

\subsection{Fine-Grained Workload Representation}\label{sec:Contribution:ProgrammingModel}

\insertFigure{ProgrammingModel}{1}{0}{0}{
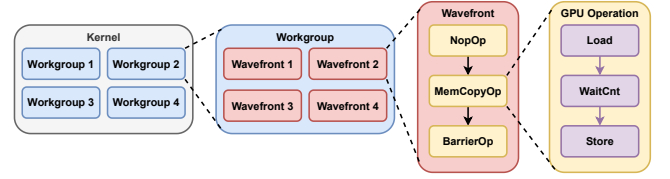
Abstract view of a GPU kernel broken down into fine-grained \loadinst{}-\storeinst{} granularity in \ours{}.
}

We propose to model the GPU device at the \loadinst{}-\storeinst{} granularity.
The critical operations in distributed ML are the control and data paths over the on-chip and inter-GPU networks.
They can effectively be modeled by capturing the fine-grained \loadinst{}-\storeinst{} instructions.
Simulating the exact GPU binary at the finest granularity contains unnecessarily detailed information and hurts simulation scalability.

A high-level overview of the GPU kernel decomposed in \ours{} is shown in~\autoref{fig:ProgrammingModel}.
It captures the execution in the manner in which the GPU programming model does: a workgroup comprises multiple wavefronts, each running multiple GPU operations in sequence, where each GPU operation itself is ultimately a sequence of \loadinst{}-\storeinst{} primitives. 
Below, we articulate the fine-grained workload representation in \ours{} in a bottom-up approach.

\subsubsection{GPU Instructions}

As we proposed, we define primitive GPU instructions at the \loadinst{}-\storeinst{} granularity.
These GPU instructions become the unit of simulation in \ours{}.
The GPU instructions can be grouped into three: (\rom{1}) data instructions, (\rom{2}) control instructions, and (\rom{3}) others.
For these \loadinst{}-\storeinst{}, the source or destination memory location may reside either locally or remotely.

\paratitle{Data Instructions}
They represent the movement of data between the CU and memory, as defined below:
\begin{itemize}
    \item \loadinst{}: A CU loads a chunk of data from memory to the register file.
    \item \storeinst{}: A CU stores a chunk of data from the register file to memory.
\end{itemize}

\paratitle{Control Instructions}
These instructions also create load and store requests.
However, we differentiate these from the data operations to emphasize their nature as control path instructions.
They are listed below:
\begin{itemize}
    \item \acquireinst{}: A CU loads a semaphore value and checks if it is released by another CU.
    \item \releaseinst{}: A CU stores a semaphore value to mark that it is released and can be acquired by another CU.
\end{itemize}

\paratitle{Other Instructions}
Finally, we add two more primitive instructions, as below, to capture non-memory executions.
\begin{itemize}
    \item \reduceinst{}: Abstracts all other arithmetic or logical operations taking place in the CU.
    \item \waitcntinst{}: Puts the CU on wait until the number of in-flight \loadinst{} and \storeinst{} drops to a specific threshold, to control intra-wavefront memory dependencies.
\end{itemize}

\subsubsection{GPU Operations}\label{sec:Contribution:GpuOperations}

While these primitive \loadinst{}-\storeinst{} instructions capture the fine-grained behaviors of a GPU, they are too fine-grained to capture the logical operation the programmer leverages.
To mitigate this gap, we introduce GPU operations.
A GPU operation is a sequence of primitive GPU instructions, and it denotes a meaningful, functional programming unit.
Examples include loading data from a range of memory addresses or synchronizing all threads within a workgroup.

For the scope of this work, \ours{} implements several data- and control-related GPU operations.
However, any new GPU operation can be easily implemented, as it is simply a sequence of multiple \loadinst{}-\storeinst{} GPU instructions.

\paratitle{Data Operations}
These operations are used to load and store a range of data.
We predefine three data operations as below:
\begin{itemize}
    \item \loadop{}: {
        A wrapper of the \loadinst{} instruction.
        Loads a range of data from memory to the CU.
    }
    \item \storeop{}: {
        A wrapper of the \storeinst{} instruction.
        Stores a range of data from the CU to memory.
    }
    \item \memcpyop{}: {
        Represents memory-to-memory copy operations.
        Data is first loaded to the CU by the \loadinst{} instruction, followed by the \waitcntinst{} instruction to enforce a memory fence.
        Finally, the \storeinst{} instruction writes the loaded data from the CU to the destination memory space.
    }
\end{itemize}

\paratitle{Control Operations}
These two control operations wrap the semaphore instructions, as shown below:
\begin{itemize}
    \item \acquireop{}: Issues a \acquireinst{} instruction to acquire the semaphore.
    \item \releaseop{}: Issues a \releaseinst{} instruction to release the semaphore.
\end{itemize}

\paratitle{Other Operations}
We define \reduceop{} to wrap the \reduceinst{} instruction to capture all non-memory arithmetic and logical operations.
Interestingly, we note that dispatching no GPU instructions may itself be a meaningful operation.
An example is an operation to halt execution until a certain condition is met.
We define two such operations for GPU execution control without any memory instructions.
\begin{itemize}
    \item \reduceop{}: {
        A wrapper of the \reduceinst{} instruction.
        Represents all arithmetic and logical operations without memory operations.
    }
    \item \nopop{}: {
        This operation halts the execution of a workgroup until all its wavefronts reach this operation (i.e.,~\texttt{\_\_syncthread} operation). 
    }
    \item \barrierop{}: {
        While \nopop{} enforces intra-workgroup synchronization of wavefronts, \barrierop{} enforces inter-workgroup synchronization.
    }
\end{itemize}

\subsubsection{Workgroup and Wavefront}

Now that we have defined the logical execution operations a GPU can execute, defining a workgroup and wavefront is straightforward.
A workgroup is simply a sequence of GPU operations, executed over a single CU.
A workgroup comprises multiple wavefronts, a unit of lock-step execution in the CU.
For data operations, all wavefronts execute the \loadinst{} and \storeinst{} instructions, simulating each wavefront processing different ranges of memory addresses.
For control path operations, on the other hand, we implement only wavefront zero to issue the \loadinst{} and \storeinst{} instructions, assuming a control message (i.e., reading a semaphore value) is single cache-line-sized.

\subsubsection{Kernel}

Finally, a kernel is a set of workgroups.
When a kernel is dispatched to a GPU, each workgroup is mapped to an individual CU and executes in parallel.

To summarize, as in~\autoref{fig:ProgrammingModel},
(\rom{1})~a GPU instruction is a primitive \loadinst{}-\storeinst{} instruction,
(\rom{2})~a GPU operation is a sequence of GPU instructions, denoting a logically meaningful execution unit,
(\rom{3})~a workgroup comprises multiple GPU operations in sequence, organized in multiple wavefronts, and
(\rom{4})~a kernel is a group of workgroups that run in parallel on multiple CUs.

\subsection{\mscclpp{} Custom Collective Support}\label{sec:Contribution:customcollsupport}

Now that \ours{} represents the workload in a fine-grained representation, this easily enables the simulation of custom collective algorithms leveraging the \mscclpp{} representation.
As explained in~\autoref{sec:Background:mscclpp}, \mscclpp{} JSON representation captures arbitrary collective algorithms in per-GPU, per-workgroup operations such as \putop{} or \copyop{}.
Therefore, we simply implement a straightforward translator to represent \mscclpp{} operations in \ours{} GPU operations.
For example, a \mscclpp{} \putop{} operation (writing chunks of data from a local GPU to a remote GPU) is translated into \memcpyop{}.
The same applies to the \getop{} and \copyop{} operations.
\mscclpp{} inter-GPU control operations, \signalop{} and \waitop{}, are parsed into \releaseop{} and \acquireop{} operations, respectively.

\subsection{End-to-End Workload Execution}

\insertFigure{KernelExecution}{1}{0}{0}{
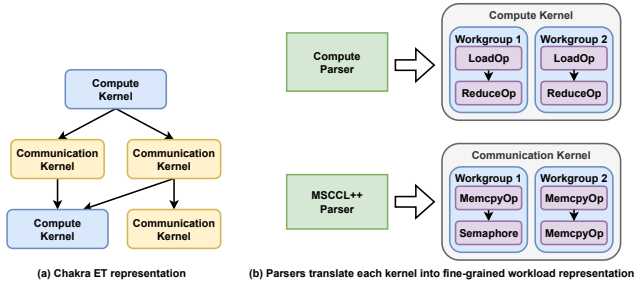
End-to-end workload simulation flow through \ours{}.
}

\ours{} inherits the end-to-end workload simulation from \astrasimtwo{} via MLCommons Chakra ET adoption.
As shown in~\autoref{fig:KernelExecution}, Chakra ET captures workloads at kernel granularity: a trace is a composition of compute and communication kernels and the dependencies between them.
\astrasimtwo{} dispatches coarse-grained events through parsing Chakra ET.
Instead, \ours{} implements parsers to decompose such kernels into fine-grained workload representations.
For communication kernels, we leverage the \mscclpp{} parser discussed in~\autoref{sec:Contribution:customcollsupport}.
Likewise, compute kernels can be decomposed into multiple workgroups, each executing \reduceop{} operations.
\astrasimtwo{} used very basic hardware models to simulate resource contention, such as dispatching one kernel at a time.
However, in \ours{}, all kernels are decomposed into the common, fine-grained representation and simulated by a single model.
Therefore, resource contention between multiple compute or communication kernels is naturally captured without such arbitrary restrictions.

\subsection{GPU Model}

So far, we have discussed how \ours{} represents the target workload in a fine-grained, \loadinst{}-\storeinst{} instruction representation.
\ours{} implements a new execution model to simulate this fine-grained representation in detail to capture the actual program execution.
Below, we explain this execution model in a top-down approach.

\subsubsection{GPU Model}

As the name suggests, the \gpumodel{} abstracts a physical GPU and consists of multiple CUs.
The \ours{} simulation starts by dispatching a kernel (parsed in the fine-grained representation) to \gpumodel{}.
The \gpumodel{} iterates over the workgroups and maps each workgroup to a (free) CU in a round-robin order, thereby modeling CU resource conflicts.
The number of workgroups that each CU can accommodate can be tuned by the user.

\subsubsection{Compute Unit}

When a workgroup is dispatched to a CU, it executes the wavefronts from the workgroup.
The CU chooses a wavefront that is ready (i.e., not stalled) and executes the GPU operations in sequence.
During the execution, if a wavefront stalls due to control path operations, the CU executes another ready wavefront, modeling wavefront-level parallelism.
If all GPU operations of a wavefront are finished, then the wavefront is marked as done.
When all wavefronts of a workgroup are completed, the workgroup is marked as finished and retires.

The items below explain how the non-memory-based GPU operations execute in the CU:
\begin{itemize}
    \item \nopop{}: {
    As mentioned in~\autoref{sec:Contribution:GpuOperations}, this operation does not have any GPU instructions.
    Instead, when executing this operation, the CU simply puts the wavefront in a stalled state, and checks if all other wavefronts in the workgroup are also stalled.
    If so, \nopop{} completes and all wavefronts are marked as ready to execute the next operation.
    }
    \item \barrierop{}: The execution follows the same logic as \nopop{}, but enforces inter-workgroup synchronization rather than intra-workgroup synchronization.
    \item \reduceop{}: Occupies the CU for a certain number of simulation cycles to mimic the arithmetic/logical operations, then retires the operation.
\end{itemize}

Other GPU operations comprise \loadinst{} or \storeinst{} instructions.
For example, \loadop{} requires a \loadinst{} instruction to load a value from memory.
Likewise, \acquireop{} wraps a \acquireinst{} instruction, which dispatches a \loadinst{} instruction to check the semaphore value.
When executing these operations, the CU injects cache-line-sized network requests, namely \wfreq{}s, into the network simulator in each cycle.

\subsubsection{Wavefront Request}

A \wfreq{} is a single cache-line-sized (e.g., 128\,B) transfer that a CU dispatches to the network.
In \ours{}, the \wfreq{} is the unit transfer that the network backend simulates.
The network layer receives these \wfreq{}s and simulates the local (on-chip) and remote (scale-up and scale-out) network traffic.
When a CU executes the GPU operation of a wavefront, it generates multiple \wfreq{}s depending on the operation.
\begin{itemize}
    \item \acquireop{} and \releaseop{}: {
    Acquiring and releasing a semaphore requires reading and writing the semaphore value, respectively.
    As we assume each semaphore fits in a single cache line, executing these operations generates one \wfreq{} from wavefront zero.
    }
    \item \loadop{}, \storeop{}, and \memcpyop{}: {
    They generate multiple \wfreq{}s to read or write a memory span.
    }
\end{itemize}

\subsubsection{Loop Unrolling}

\insertFigure{LoopUnrolling}{1}{0}{0}{
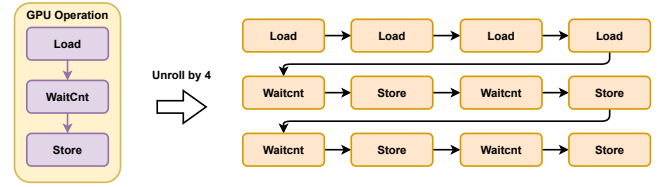
A \memcpyop{} operation unrolled by four times.
}

When a CU executes a wavefront, it can dispatch one cache-line-sized \wfreq{} in each cycle.
A GPU may facilitate intra-wavefront, instruction-level parallelization by dispatching multiple in-flight memory requests simultaneously.
In \ours{}, we implement this via tunable loop unrolling.
\autoref{fig:LoopUnrolling} exemplifies the unrolling of a \memcpyop{} operation.
\memcpyop{} requires the repetition of
(\rom{1})~loading cache-line-sized data,
(\rom{2})~making sure the data has been loaded by the \waitcntinst{} instruction, then
(\rom{3})~storing the data to the destination.
Instead of repeating the three primitive instructions for each cache line, \autoref{fig:LoopUnrolling} unrolls it by a factor of four.
The unrolling enables overlapping four \loadinst{} \wfreq{}s, capturing intra-wavefront parallelism.

\subsection{Detailed Simple Network Modeling}

With the new custom collective and GPU modeling capabilities, \ours{} is equipped with detailed, fine-grained simulation opportunities with NoC-level requests and cache-line-sized transfers.
To better accommodate this opportunity, we upgrade the \simplenetwork{} network simulation backend from \astrasimtwo{} to simulate NoC-level details alongside the inter-GPU network.
Note that \ours{} still supports other network simulation backends with detailed NoC models through the network API.

The high-level insight is depicted in~\autoref{fig:SimpleNetworkModel}.
The old \simplenetwork{} network model, shown in~\autoref{fig:SimpleNetworkModel}(a), used each GPU as the building block.
The network representation was restricted to simple GPU connections and simulated through coarse-grained transfers.
The updated \simplenetwork{} network is shown in~\autoref{fig:SimpleNetworkModel}(b).
Now the target network topology is represented with NoC-level details, such as CUs, HBM channels, and I/O ports.
As the \gpumodel{} of the \ours{} infrastructure dispatches fine-grained \wfreq{}, the updated \simplenetwork{} with the NoC-level representation is able to simulate both local and remote operations.
Memory channels are added to \simplenetwork{} to model the NoC and memory traffic.
The network representation of \simplenetwork{} uses a hierarchical file-based representation and modular routing function to allow scalability and quick interchanging of network topologies or routing.

\insertFigure{SimpleNetworkModel}{1}{0}{0}{
Updated \simplenetwork{} network simulation backend.
}

\subsection{Infrastructure as a Graph}
We present \infragraph{}, a standard, portable representation for AI and high-performance computing~(HPC) network infrastructure.
\infragraph{} formalizes infrastructure topology as a directed, attributed graph in which vertices represent hardware like GPUs, network interface controllers~(NICs), and storage, while edges represent the connections between the hardware components.
Property annotations such as bandwidth enhance the expressiveness of the graph.
This representation is based on the principle that complex infrastructure topologies can be naturally expressed through graph constructs.

Users define a graph by describing reusable infrastructure objects and topology relationships and programmatically expanding them into a complete graph representation.
Reusing components allows a compact description to expressively represent a fully expanded graph, while programmatic graph construction enables automatic workflows.

\subsubsection{Device Description}\label{subsubsec:schema}

We first define the primitives used to describe a \deviceinst{}:

\begin{itemize}
    \item \componentinst{}: Hardware unit within a device such as a CPU, GPU, or peripheral component interconnect express~(PCIe) bridge. 
    \item \linkinst{}: Named connection container with physical properties such as bandwidth or latency.
\end{itemize}

\subsubsection{\infragraph{} Graph Construction}
\label{subsubsec:construct}

A core design principle of \infragraph{} is to make the definition compact by reusing modular definitions.
Rather than manually defining every endpoint and interconnection in a large-scale AI cluster, users describe reusable infrastructure objects that can be programmatically expanded into a complete graph.
For example, in a scale-out topology across multiple host servers, a single host can be represented as a \deviceinst{} where hardware \componentinst{} such as a CPU, GPU, or PCIe bridge are represented as vertices and PCIe \edgeinst{}s are the edges.
Instead of repeatedly listing the \componentinst{}s and \edgeinst{}s multiple times for all hosts, a user can elect to programmatically construct them. 
The below primitives enable large construction:
\begin{itemize}
    \item \deviceinst{}: Subgraph template for device hardware, containing \componentinst{} and \edgeinst{}s.
    \item \instanceinst{}: Device instantiation alias.
    \item \infrainst{}: Top level graph container.
    \item \deviceinst{}.\edgeinst{}: Edge of an infrastructure graph defined by two \deviceinst{} endpoints and a connecting \linkinst{}.
\end{itemize}

\subsubsection{Blueprints}
We also offer pre-built, composable templates for common hardware platforms and network fabrics.
Device blueprints define the internal hardware structure of a single platform---components, links, and intra-device edges---and serve as reusable building blocks.
Fabric blueprints compose device instances into full network topologies.
These blueprints are parameterized (e.g., the number of hosts) and automatically instantiated.
For example, \texttt{SingleTierFabric} fabric blueprint implements a flat single-switch-layer topology for small-scale deployments, while \texttt{ClosFatTreeFabric} produces a scalable hierarchical topology parameterized by switch port count and network depth, automatically computing switch counts and wiring all links per the standard CLOS construction.

\insertFigure{infragraph_topo}{1}{0}{0}{
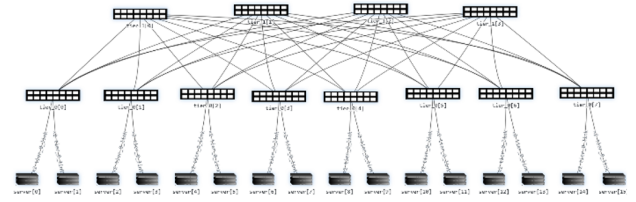
Clos fabric generated and visualized using \infragraph{} blueprint and visualizer.
}

\subsection{InfraGraph Toolchain}

\subsubsection{Translator}
We build a translator that takes the \infragraph{} description and automatically translates this into a physical system description that each network backend in \astrasim{} can understand.
The translator has three backend translators for the publicly available network backends of \astrasim{}: \simple{}, \ns{}, and \htsim{}.
Each backend translator automatically derives its required parameters from the annotated graph.
The \htsim{} translator infers fat-tree structural parameters from the topology; \ns{} assigns identifiers to GPU ranks, NICs, and switches and reads per-link properties from edge annotations; and the \simple{} translator additionally detects topology patterns to decompose large node counts into multi-dimensional groups for collective communication modeling.
In all cases, the same \infragraph{} description produces valid configurations across all backends, enabling direct cross-backend comparison under identical infrastructure assumptions.

\subsubsection{Visualizer}
The visualizer receives an \infragraph{} and automatically generates network connectivity plots.
This allows users to see whether the network graph they defined is what they indeed intended.

\subsubsection{Fully Qualified Graph and \infragraph{} Service}
\infragraph{} expands a topology description into a fully qualified graph composed of infrastructure nodes and edges.
Each device, component, port, and link is represented using unique hierarchical identifiers, enabling accurate topology representation, communication path discovery, graph traversal, connectivity analysis, and topology-aware simulation across large-scale AI and HPC infrastructures.
An infrastructure node represents the fundamental entity within the infrastructure graph and corresponds to a specific endpoint in the system topology.
A node follows the naming convention: \texttt{<device-instance>.<index>.<component>.<index>}.
This hierarchical naming structure enables precise identification of components within large-scale heterogeneous systems.
An infrastructure edge represents a communication relationship between two infrastructure nodes.
Edges define connectivity within the graph and model communication links between components.
An example of an edge representation for a switch is: \texttt{(switch.0.asic.0, switch.0.port.0, pcie)} where it is represented as a source node, a destination node, and the link connecting both nodes.

\insertFigure{GetPutComparison}{1}{0}{0}{
Simulated collective performance of \getop{}- and \putop{}-based \reducescatter{} with 32 GPUs.
}

\section{Case Studies}\label{sec:Results}

In this section, we run various case studies to showcase how \ours{} enables new design space exploration opportunities that previous simulators could not capture.
We wish to highlight that this section aims to demonstrate various use cases and the applicability of the \ours{} simulation infrastructure, not necessarily showcasing quantitative optimization measures of specific products or infrastructures.
Therefore, we evaluate a generic GPU architecture with power-of-two configuration values.

\insertFigure{GetPutAllGather}{1}{0}{0}{
Simulated collective bandwidth of \getop{}- and \putop{}-based \allgather{} with 16 GPUs, with and without arbitration between control and data messages.
}

\subsection{Target GPU Architecture}

For the fine-grained GPU modeling case studies, we implement a generic GPU architecture.
We model a two-dimensional mesh NoC with 32 routers (8$\times$4) with 1\,TiB/s on-chip links, each equipped with four CUs (i.e., 128 CUs total per GPU).
The topmost and bottommost NoC routers are also connected to 16 memory channels each, providing 4\,TiB/s cumulative memory bandwidth.
Likewise, the leftmost and rightmost NoC routers are equipped with four I/O ports each, driving 1\,TiB/s cumulative scale-up bandwidth with 1\,$\mu$s link latency per GPU.
In total, each GPU is modeled as a collection of 448 endpoints.

\subsection{Collective Algorithm Design}

We first demonstrate how \ours{} enables a new realm of modeling collective algorithm designs and assists network designers in optimizing the target system, through the fine-grained, faithful control path modeling.
Unlike existing infrastructures that don't support custom collective algorithms, \ours{} enables the comparison of using \getop{} versus \putop{} for data transfers, which imply different synchronization requirements.

We evaluate the bandwidth of \getop{}- and \putop{}-based \reducescatter{} algorithms over a 32-GPU cluster with 32 workgroups per GPU.
Simulated collective bandwidths (i.e., buffer size divided by collective time) are shown in~\autoref{fig:GetPutComparison}.
The results show that \getop{}-based \reducescatter{} outperforms \putop{}-based \reducescatter{} for large collectives.

\paratitle{Modeling Insight}
For \putop{} operations, the sender must notify the receiver after the completion of the transfer (i.e., \releaseop{} should follow after a remote \storeinst{}).
This incurs unnecessary control traffic and hinders the overlap of data transfer and reduction computation, as the receiver must run \acquireop{} before \reduceop{}.
However, \getop{} operations eliminate synchronization.
Once the initiator receives the data via a remote \loadinst{}, that GPU can immediately start the reduction.
This enables compute-communication overlap at cache-line granularity.

We also compare \getop{}- versus \putop{}-based collective algorithms for \allgather{}, using 16 GPUs with 60 workgroups each.
\autoref{fig:GetPutAllGather} captures the result.
Notably, the \getop{}-based collective is less performant than the \putop{}-based one.

\paratitle{Modeling Insight}
Unlike \reducescatter{}, \allgather{} does not have a reduction operation, eliminating the benefits of \getop{} operations.
Rather, as the buffer size becomes large, the control messages (i.e., requests to the remote GPU to send data) are being blocked by data responses.
\putop{} operations, on the other hand, push the data first to the network, and the responses (i.e., acknowledgment messages after remote \storeinst{}) have a marginal effect on performance even if blocked.
By using fair arbitration of control and data messages, we minimized the performance gap.

Different collective algorithms, even as small as using \getop{} versus \putop{}, have a significant impact and implications for performance and network design requirements and configurations.
They exemplify the necessity of having fine-grained control path modeling that \ours{} enables.

\subsection{GPU Architecture Exploration}

\insertFigure{LoopUnrollingResult}{1}{0}{0}{
Simulated \alltoall{} performance of varying loop unrolling factors with 16 GPUs.
}

\ours{} fine-grained modeling is also applicable to GPU architecture design and enables new design space exploration capabilities.
For example, \ours{} can be used to model and determine the degree of required intra-wavefront, instruction-level parallelism to optimize collective performance.
\autoref{fig:LoopUnrollingResult} measures the \alltoall{} performance using 16 GPUs, 60 workgroups per GPU.

\paratitle{Modeling Insight}
The result indicates that increased instruction-level parallelism, which enables each CU to dispatch multiple concurrent \wfreq{}, is beneficial for improving bandwidth-bound collectives.
Notably, (\rom{1})~the benefit saturates after hitting the maximum number of \wfreq{} each CU can dispatch at a time, and (\rom{2}) instruction-level parallelism is not a relevant optimization for latency-bound, small-sized collectives.

\insertFigure{WFReqExploration}{1}{0}{0}{
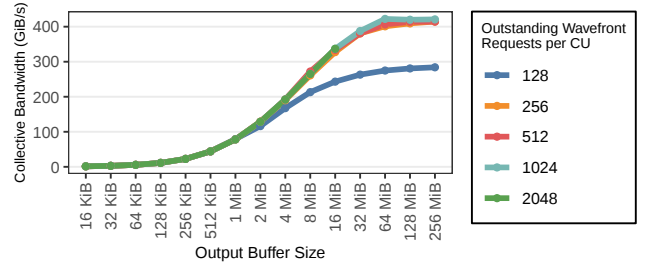
Simulated \allgather{} bandwidth with 32 GPUs, with different numbers of maximum outstanding \wfreq{} limits per CU.
}

We also evaluate how the maximum number of \wfreq{} each CU can dispatch impacts the collective bandwidth.
\autoref{fig:WFReqExploration} summarizes the \allgather{} performance with 32 GPUs, 62 workgroups per CU.

\paratitle{Modeling Insight}
This configuration is a proxy for the register file size, as it is the determining factor of how many cache-line-sized requests a CU can service concurrently.
Similar to the loop unrolling observation, (\rom{1})~register file size does not have a meaningful impact on control-path-dominated, latency-bound collectives, and (\rom{2})~the benefit saturates quickly after hitting a specific register file size.

These examples highlight the effectiveness of having fine-grained, \loadinst{}-\storeinst{}-level architecture modeling that enables new architectural explorations.

\subsection{Fine-Grained Simulation Scalability}

\insertFigure{Scalability}{1}{0}{0}{
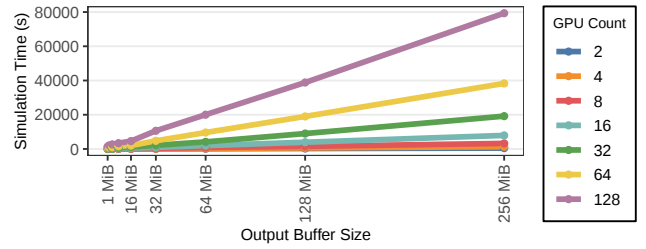
Wall-clock simulation time of \allgather{} for 1--256\,MiB output buffer sizes, when modeling 2--128 GPUs.
}

We also analyze the scalability of the fine-grained, cache-line-sized \loadinst{}-\storeinst{} modeling with the updated \simple{} network's NoC-level simulation.
Specifically, we measure the wall-clock simulation time of \allgather{} collective communication with 32 workgroups per GPU, for 1--256\,MiB output buffer sizes.
We scale the target system to 2--128 GPUs.
Notably, a 128-GPU cluster simulates 57,344~(128 GPUs$\times$448) endpoints simultaneously at 128\,B granularity.
The results are summarized in~\autoref{fig:Scalability}.
We also plot the simulation throughput (i.e., how many nanoseconds \ours{} can simulate per wall-clock second) in~\autoref{fig:ScalabilityThroughput}.

\paratitle{Simulation Insight}
The larger the collective size, the larger the number of 128\,B-sized \wfreq{}s the simulation requires to be simulated.
Therefore, for all scenarios, the simulation time is linear in the output buffer size.
Regardless of the number of \wfreq{}s dispatched to the network simulator, the simulator retained similar simulation throughput.
Rather, it is determined by the target system scale being modeled.

\insertFigure{ScalabilityThroughput}{1}{0}{0}{
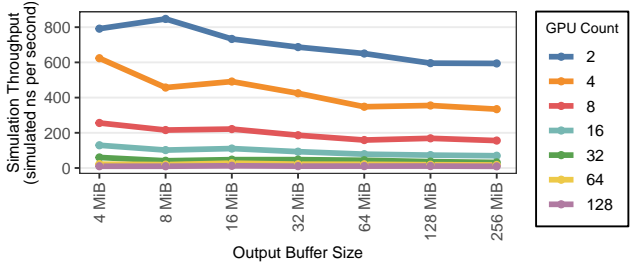
Simulation throughput (i.e., simulated nanoseconds per wall-clock second) for 2--128 GPUs.
}

\subsection{Scale-Out Infrastructure Simulation}

Finally, we validate \ours{}'s ability to capture distinct networks via a standardized \infragraph{}.
We simulate a 1\,MB ring \allreduce{} across eight GPUs interconnected via a Clos fabric shown in~\autoref{fig:infragraph_topo}
using the \ns{} packet-level backend.
The system is instantiated from the \infragraph{} device blueprint, with per-device and per-link properties bound via the annotation layer.
Results are summarized in \autoref{tab:infragraph_dgx}.
The simulation properly modeled the Clos scale-out network, yielding a minimum flow completion time~(FCT) of 11,250 ns and a standalone FCT of 11,857 ns.
The infrastructure does not incur any packet drops, confirming that the ring traffic pattern remains lossless within the simulated fabric.

\section{Related Work}\label{sec:RelatedWork}

We largely group previous simulators into two domains: (\rom{1})~GPU binary simulators and (\rom{2})~distributed ML simulators.

\paratitle{GPU Binary Simulators}
GPGPU-sim~\cite{gpgpu-simGPGPUSim} and MacSim~\cite{macsim} are cycle-accurate simulators of GPU executables, modeling the complete instruction set architecture~(ISA) of a GPU.
Accel-sim~\cite{khairy2020accelsim} is another example of an instruction-level GPU binary simulator.
We note that such fine-grained, cycle-accurate or instruction-level simulators take a long time to simulate a single GPU, making them prohibitive for multi-GPU simulations.
Also, ISA-level simulation is not relevant to distributed ML as the communication performance is dominated by network operations.

\paratitle{Distributed ML Simulators}
Multiple distributed ML simulators have been introduced to the research community.
ATLAHS~\cite{shen2025atlahs} models AI use cases, though it mainly looks at high-performance computing~(HPC) or storage workloads.
Calculon~\cite{isaev2023calculon} provides detailed analytical modeling for LLM system co-design, and TrioSim~\cite{li2025triosim} extrapolates single-GPU operator traces to large-scale multi-dimensional parallel execution.
Maya~\cite{yarlagadda2026maya} and Phanthora~\cite{qin2025phantora} reuse model code to easily capture and model workload information.
SimAI~\cite{wang2025simai} leverages parallelized discrete event simulation to improve scalability.
Arcadia~\cite{2023arcadia} scales further to support multi-job scheduling scenarios.
Notably, these simulators don't support arbitrarily customized collective algorithms and default to textbook algorithms.
Also, none of the simulators capture CU-level, cache-line-sized GPU behaviors, crucial to modeling important control-path latencies.

\begin{table}[t]
\centering
\caption{\ns{} \allreduce{} simulation result.}
\label{tab:infragraph_dgx}
\begin{tabular}{c|c}
\hline
\textbf{Metric} & \textbf{Clos Network} \\
\hline
AllReduce Completion Time ($\mu$s) & 165.98 \\
\hline
Achieved Bus Bandwidth (Gbps) & 88.45 \\
\hline
Min FCT (ns) & 11{,}250 \\
\hline
Max FCT (ns) & 14{,}552 \\
\hline
Avg FCT (ns) & 11{,}477 \\
\hline
Standalone FCT (ns) & 11{,}857 \\
\hline
Peak FCT Overhead (ns) & 2{,}695 \\
\hline
\end{tabular}
\end{table}

\section{Conclusion}\label{sec:Conclusion}

We introduce \ours{}, an update to the \astrasim{} framework equipped with high-fidelity modeling capabilities.
\ours{} enables new design space exploration opportunities by capturing fine-grained control path and architectural details with custom collective modeling.
\ours{} also introduces \infragraph{}, a common representation to capture the exact distributed ML infrastructure.

\section*{Acknowledgments}

We express our deepest gratitude to the original developers and authors of the \astrasim{} infrastructure: Saeed Rashidi, Srinivas Sridharan, Sudarshan Srinivasan, and Taekyung Heo.

AMD, the AMD Arrow logo, and combinations thereof are trademarks of Advanced Micro Devices, Inc.
Other product names used in this publication are for identification purposes only and may be trademarks of their respective companies.

\paratitle{AI Usage}
Generative AI models (i.e., LLMs) are only used to check the writing quality and grammatical errors of the paper.
All contents are solely created by the authors.

\bibliographystyle{ACM-Reference-Format}
\bibliography{reference}

\end{document}